\documentclass[twoside,fleqn]{ActaStyle}
\usepackage{times,cite}

\usepackage{overpic}
\usepackage{axodraw}
\usepackage{graphicx,epsfig}    

\newcommand{\ba}{\begin{eqnarray}}
\newcommand{\ea}{\end{eqnarray}}
\newcommand{\dsp}{\displaystyle}

\def\authorheading{J.~Bijnens}

\def\shorttitle{Decays of $\eta$ and $\eta'$ and what can we learn from them?}

\renewcommand{\pagerange}[2]{%
\newcount{\prvastrana}\prvastrana=#1
\newcount{\poslednastrana}\poslednastrana=#2
\setcounter{page}{#1}
\begin{flushright}
LU TP 05-42\\
hep-ph/0511076\\
revised february 2006
\end{flushright}
}
\renewcommand{\publ}{\begin{minipage}{0.9\textwidth}
Talk presented at the Eta'05 Workshop on production and decay
of $\eta$ and $\eta'$ mesons
15th---18th September 2005, Jagellonian University, Cracow, Poland
\end{minipage}
}

\begin{document}

\newpage

\pagerange{1}{12}

\title{%
DECAYS OF $\eta$ AND $\eta'$ AND WHAT CAN WE LEARN FROM THEM?
}

\author{
Johan Bijnens\email{bijnens@thep.lu.se}$^*$
}
{
$^*$Department of Theoretical Physics, Lund University,\\
S\"olvegatan 14A, SE 223 62 Lund, Sweden
}

\day{November 7, 2005}

\abstract{%
In this talk a short overview of $\eta$ and $\eta'$ decays is given with an
emphasis on what can be learned for the strong interaction from them.
The talk consists of a short introduction to Chiral Perturbation
Theory, a discussion of
$\eta\to3\pi$ beyond $p^4$ and some of the physics involved in 
$\eta'\to\eta\pi\pi,3\pi$ as well as an overview of anomalous processes.
}


\section{Introduction}
\label{sec:intr} \setcounter{section}{1}\setcounter{equation}{0}

This talk gives an introduction to some of the basic strong interaction
issues we face in $\eta$ and $\eta'$ decays. It will not cover weak decays.
Simple dimensional analysis leads to branching ratios below $10^{-11}$ for
weak $\eta$-decays and below $10^{-12}$ for weak $\eta'$ decays. These will
obviously not be observed in the near future. It is possible to construct
models that enhance $\eta$ and $\eta'$ decays due to 
physics beyond the standard
model to observable rates but these models tend to be ugly in order
to avoid the very stringent constraints from kaon decays and other sources,
some examples can be found in~\cite{etahandbook}.

Let me remind you of the $\eta$-handbook~\cite{etahandbook} where you can find
a series of lectures on the basis of $\eta$ and $\eta'$ physics.
The theory lectures relevant for this talk in there are those by
Kroll~\cite{Kroll}, Bijnens and Gasser~\cite{BG}, Ametller~\cite{Lluis},
Holstein~\cite{Holstein}, Shore~\cite{Shore} and Bass~\cite{Bass}.
There are also a set of related theory talks in this conference,
those by Bass~\cite{Bass2},
Borasoy~\cite{Borasoy}, 
Oset~\cite{Oset} and Martemyanov~\cite{Martemyanov},
while most of the other talks today are the
experimental talks related to $\eta$ and $\eta'$ decays at KLOE and WASA.

\section{Pseudoscalars are special}

The Lagrangian of Quantum Chromodynamics (QCD) is obviously
invariant under the interchange of the three light quarks, $u$, $d$ and $s$,
if they have equal mass. This leads to the vector symmetry $U(3)_V$.
But
\be
{\cal L}_{QCD} =  \sum_{q=u,d,s}
\left[i \bar q_L D\hskip-1.3ex/\, q_L +i \bar q_R D\hskip-1.3ex/\, q_R
- m_q\left(\bar q_R q_L + \bar q_L q_R \right)
\right]\,.
\ee
Here $q_L$ and $q_R$ are respectively the left and right handed quark spinors,
$D\hskip-1.3ex/\,= \gamma^\mu D_\mu$ is the contraction of the covariant
derivative including the gluon field with the Dirac gamma matrices
and $m_q$ is the (current) quark mass.
So, if $m_q = 0$, we have an enlarged (chiral) symmetry:
$U(3)_L\times U(3)_R$, where the left and right handed particles can be
rotated into
each other independently. 

But hadrons do not come in parity doublets: the chiral symmetry must be broken
in the real world. There exists also a
 few (very) light hadrons: $\pi^0\pi^+\pi^-$ and $K,\eta$.
The existence of the latter as well as the nonexistence of the
degenerate parity doublets
can be understood from spontaneous Chiral Symmetry Breaking.

\begin{figure}[t]
\begin{minipage}[t]{0.48\textwidth}
\begin{center}
\begin{overpic}[width=\textwidth]{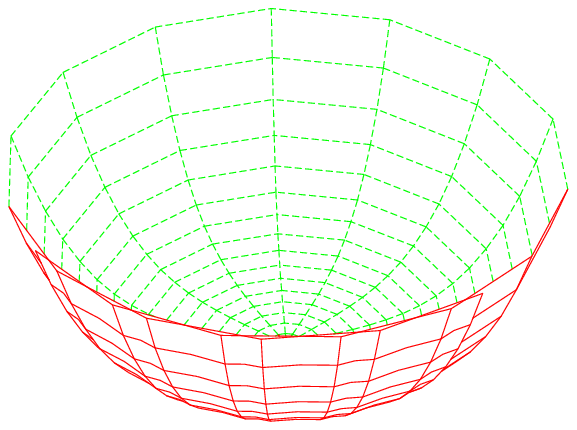}
\end{overpic}
\end{center}
\caption{The potential $V(\phi)$ for an unbroken symmetry.}
\label{figunbroken}
\end{minipage}
\begin{minipage}[t]{0.48\linewidth}
\begin{center}
\begin{overpic}[width=\textwidth,unit=0.5pt]{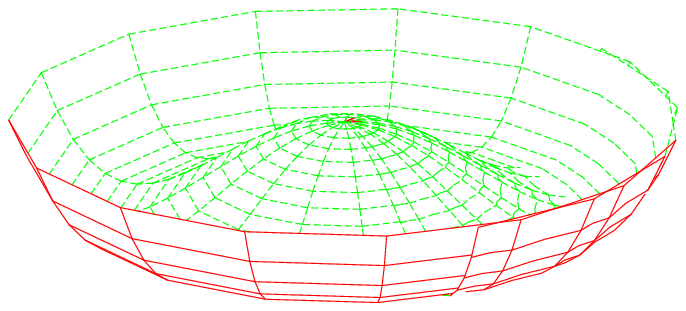}
\SetScale{0.5}
\SetWidth{3.}
\LongArrow(150,-30)(290,30)
\end{overpic}
\end{center}
\caption{The potential $V(\phi)$ for a spontaneously broken symmetry.
The arrow indicates the choice of vacuum.}
\label{figbroken}
\end{minipage}
\end{figure}

Let us first discuss spontaneous symmetry breaking for a simple
$U(1)$ symmetry for a complex scalar field $\phi(x)\to e^{i\alpha}\phi(x)$.
This means that when plotting the potential $V(\phi)$ along the $z$-axis
as a function of the real and imaginary part of $\phi$ along the $x$ and $y$
axis, it should be symmetric for rotations around the $z$-axis.
A potential $V(\phi)$ for the unbroken case is shown in Fig.~\ref{figunbroken}.
Note that the vacuum, the lowest point is unique. All excitations around the
vacuum require climbing up the sides of the potential and are thus massive.
On the contrary, in Fig.~\ref{figbroken} we show a potential corresponding to
the $U(1)$ symmetry being spontaneously broken. The lowest energy state
is now not unique anymore. This means that due to the continuous symmetry there
is a continuum of vacua or ground states with the same energy. There exists
thus a massless mode which is the moving along the valley at the bottom
of the potential. This mode is the Goldstone Boson and can
in this case, be parameterized by the angle around the $z$-axis.
The symmetry is
spontaneously broken because we have to choose one vacuum, indicated by the
arrow. We can see that the vacuum is not invariant under the symmetry group
since the vacuum expectation value $\langle\phi\rangle\ne 0$.
There are also predictions for the interactions of this Goldstone mode.
The symmetry is still present, so the angle around the $z$-axis should not
matter. This means that only changes in the angle can contribute and
a direct consequence is that the Goldstone Bosons do not interact at low
energies. This leads to the low energy theorems. A nice review about Goldstone
Bosons and their physics is~\cite{Burgess}.

Note that the degeneracy of the set of vacuum states,
sometimes referred to as the vacuum manifold, is determined by the
group structure of the symmetry group and its broken part,
and it is on this manifold
that the Goldstone Bosons in some sense live.

For QCD, we do not have a simple field $\phi$ that has a vacuum expectation
value but it is instead $\langle\overline qq\rangle\ne0$. The chiral symmetry
group $U(3)_L\times U(3)_R$ is spontaneously broken to the diagonal or vector
subgroup $U(3)_V$. The resulting Goldstone bosons we identify with
the pseudoscalars $\pi,K,\eta$.

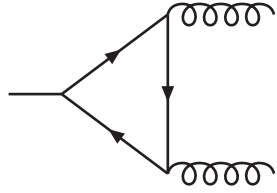
\begin{figure}
\begin{center}
\unitlength=1pt
\begin{picture}(120,90)(0,15)
\SetScale{1.0}
\SetWidth{1.}
\Line(10,50)(30,50)
\ArrowLine(30,50)(70,80)
\ArrowLine(70,80)(70,20)
\ArrowLine(70,20)(30,50)
\Gluon(70,80)(110,80){4}{4}
\Gluon(70,20)(110,20){4}{4}
\end{picture}
\end{center}
\caption{The triangle diagram which leads to a nonzero divergence for the singlet axial current.}
\label{figtriangle}
\end{figure}

This raises another problem, why is the $\eta'$ not light? This is referred
to as the $U(1)_A$-problem. The symmetry group can be decomposed into
simple groups as
\be
U(3)_L\times U(3)_R = SU(3)_L\times SU(3)_R\times
 U(1)_V\times U(1)_A\,.
\ee
The spontaneous breaking of
$SU(3)_L\times SU(3)_R$ to $SU(3)_V$
leads to 8 Goldstone Bosons and here we have $\pi,K,\eta$ as light particles
so this is fine. The $U(1)_V$ part is baryon number and is not broken.
The axial
$U(1)_A$ should not be a good symmetry of the theory in order to explain
why the $\eta'$ is heavy, even if the Lagrangian of QCD has this symmetry.
The reason is that in quantum field theory, the Lagrangian alone does not
fully specify the theory. One also needs to introduce a 
regularization/renormalization procedure. The latter is not compatible with
all global symmetries and in particular the current corresponding
to $U(1)_A$ is not conserved, this is known as the Adler-Bell-Jackiw
anomaly~\cite{ABJ}. The triangle diagram of Fig.~\ref{figtriangle}
leads to
\be
\partial_\mu A^{0\mu} = 2\sqrt{N_f}\,\omega\,,\quad\quad
\omega = \frac{1}{16\pi^2} \varepsilon^{\mu\nu\alpha\beta}
\,\mbox{tr}\,G_{\mu\nu}G_{\alpha\beta}\,.
\ee
The operator
$\omega$ consists of gluons so the nonzero divergence is strongly interacting
and it means that $U(1)_A$ is not a good symmetry for QCD. The $\eta'$
is thus allowed to be heavy.

Unfortunately, when we look at mechanisms how $\omega$ could produce the
$\eta'$ mass we see that $\omega$ is a total derivative. These can normally
be neglected in the Lagrangian so how can it have an effect? The answer was
found by 't~Hooft~\cite{instantons}. Due to special configurations with
nonzero winding number $\nu=\int d^4x\, \omega\ne0$, called instantons,
there can be an effect and it leads to a large $\eta'$ mass.

This solution lead to a new problem, the strong $CP$-problem.
One can add a term to QCD,
\be
{\cal L}_{QCD}\longrightarrow{\cal L}_{QCD}-\theta\omega,
\ee
that breaks $CP$ symmetry strongly. Experimental limits are
$|\theta|\le10^{-10}$ and we need to understand that small value.

The $\eta'$ has thus large and very interesting nonperturbative
effects and has a strong coupling to the gluon like no other hadron. Due to
the fact
that $m_s\ne\hat m =(m_u+m_d)/2$, this also affects $\eta$ physics.
This is one of the major reasons why studying $\eta$ and $\eta'$ is a very
interesting subject.

\section{Standard Chiral Perturbation Theory}

A major tool in $\eta$ decay studies is
Chiral Perturbation Theory (ChPT)~\cite{CHPT,GL1,GL2}. 
Introductions can be found
in~\cite{chptintro}. It is an effective field theory based on the Goldstone
Bosons from the spontaneous breaking of chiral symmetry as degrees of freedom.
It is an expansion in momenta and quark masses and the power counting is
really dimensional counting, called $p$-counting for a generic momentum.
The expected breakdown scale is the scale at which physics which is not
included becomes relevant. This is resonances, so the breakdown scale
is of order of the rho meson mass,
$m_\rho$, somewhat dependent on the channel one looks at.

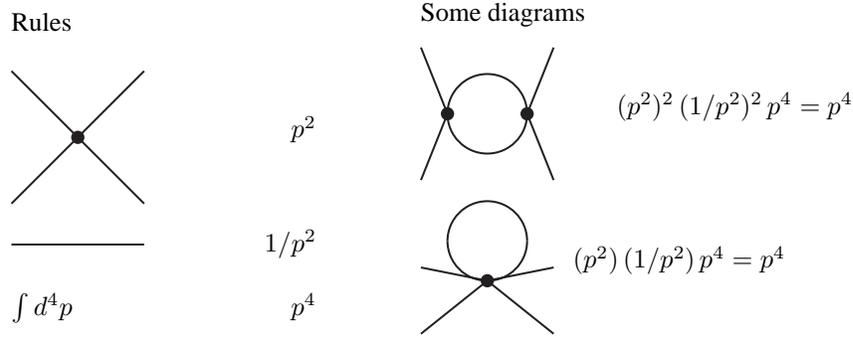
\begin{figure}
\hfill
\begin{minipage}{0.3\textwidth}
{Rules}\\[0.5cm]
\unitlength=0.5pt
\begin{picture}(100,100)
\SetScale{0.5}
\SetWidth{1.5}
\Line(0,100)(100,0)
\Line(0,0)(100,100)
\Vertex(50,50){5}
\end{picture}
\hfill\raisebox{25pt}{$p^2$}\\[0.25cm]
\unitlength=0.5pt
\begin{picture}(100,30)
\SetScale{0.5}
\SetWidth{1.5}
\Line(0,15)(100,15)
\end{picture}
\hfill\raisebox{5pt}{$1/p^2$}\\[0.25cm]
$\int d^4p$\hfill$p^4$
\end{minipage}
\hfill
\raisebox{0.cm}{
\begin{minipage}{0.50\textwidth}
{Some diagrams}\\[-1.5cm]
\begin{picture}(100,100)
\SetScale{0.5}
\SetWidth{1.5}
\Line(0,100)(20,50)
\Line(0,0)(20,50)
\Vertex(20,50){5}
\CArc(50,50)(30,0,180)
\CArc(50,50)(30,180,360)
\Vertex(80,50){5}
\Line(80,50)(100,100)
\Line(80,50)(100,0)
\end{picture}
\hskip-1cm\raisebox{25pt}
{$(p^2)^2\,(1/p^2)^2\,p^4 = p^4$}\\[0.25cm]
\unitlength=0.5pt
\begin{picture}(100,100)
\SetScale{0.5}
\SetWidth{1.5}
\Line(0,0)(50,40)
\Line(0,50)(50,40)
\CArc(50,70)(30,0,180)
\CArc(50,70)(30,180,360)
\Vertex(50,40){5}
\Line(50,40)(100,50)
\Line(50,40)(100,0)
\end{picture}
~~\raisebox{25pt}
{$(p^2)\,(1/p^2)\,p^4 = p^4$}
\end{minipage}
}
\caption{On the left hand side the power counting rules for ChPT for a
lowest order vertex, a propagator and a loop integration. 
The right side
shows how this leads to the same order for two different one-loop
diagrams.}
\label{figpipi}
\end{figure}

The fact that a power counting in momenta leads to a well defined perturbative
expansion follows from the fact that the interactions of Goldstone Bosons
vanish at zero momentum. The absence of the latter for other strongly
interacting states is why it is so difficult to build an effective theory
including resonances.
Power Counting is shown for the example of $\pi\pi$ scattering
in Fig.~\ref{figpipi}.

ChPT is a nonrenormalizable field theory. That means that, while in
principle predictive, the number of parameters increases strongly order by
order.
\begin{table}
\begin{center}
\begin{tabular}{|c|cc|cc|cc|}
\hline
      & \multicolumn{2}{c|}{ 2 flavour} & \multicolumn{2}{c|}{3 flavour} &
\multicolumn{2}{c|}{ 3+3 PQChPT} \\
\hline
$p^2$ & $F,B$ & 2 & $F_0,B_0$ & 2 &  $F_0,B_0$ &  2 \\
$p^4$ & $l_i^r,h_i^r$ & 7+3 & $L_i^r,H_i^r$ & 10+2 & 
      $\hat L_i^r,\hat H_i^r$ &  11+2 \\
$p^6$ & $c_i^r$ & 53+4 & $C_i^r$ & 90+4 &  $K_i^r$ &
       112+3\\
\hline
\end{tabular}
\end{center}
\caption{The number of free parameters in ChPT at various orders for
two and three flavours and the partially quenched case.}
\label{tabparam}
\end{table}
At lowest order there are only two parameters~\cite{Weinberg2}, at $p^4$
ten~\cite{GL2} and at $p^6$ there are 90~\cite{BCElag}.
Other cases are shown in Table~\ref{tabparam}. 

In practice
the $p^6$ coefficients $C_i^r$ are often determined from resonance
saturation depicted schematically in Fig.~\ref{figresonance}.
\begin{figure}
\begin{minipage}{0.66\textwidth}
\unitlength=0.5pt
\begin{picture}(440,100)
\SetScale{0.5}
\SetWidth{1.5}
\Line(0,100)(10,50)
\Line(0,0)(10,50)
\Text(10,10)[]{$\pi$}
\Text(10,90)[]{$\pi$}
\Vertex(10,50){5}
\Line(10,52)(100,52)
\Line(10,48)(100,48)
\Text(55,65)[]{$\rho,S$}
\Text(55,30)[]{$\rightarrow q^2$}
\Vertex(100,50){5}
\Line(100,50)(110,0)
\Line(100,50)(110,100)
\Text(120,10)[]{$\pi$}
\Text(120,90)[]{$\pi$}
\Text(220,75)[]{$|q^2|<< m_\rho^2,m_S^2$}
\Text(220,50)[]{\Large$\Longrightarrow$}
\Line(320,100)(380,50)
\Line(320,0)(380,50)
\Line(380,50)(440,100)
\Line(380,50)(440,0)
\Text(380,75)[]{$C_i^r$}
\Vertex(380,50){5}
\end{picture}
\caption{A schematic view of resonance saturation of the $C_i^r$.}
\label{figresonance}
\end{minipage}
\begin{minipage}{0.32\textwidth}
\setlength{\unitlength}{0.5pt}
\begin{picture}(190,110)
\SetScale{0.5}
\SetWidth{1}
\Line(10,50)(80,50)
\Text(8,50)[r]{$\eta$}
\Text(30,45)[t]{$p_\eta$}
\Vertex(80,50){4}
\Line(80,50)(150,100)
\Text(152,102)[l]{$\pi^+$}
\Text(180,98)[l]{$p_{\pi^+}$}
\Line(80,50)(150,50)
\Text(152,52)[l]{$\pi^-$}
\Text(180,48)[l]{$p_{\pi^-}$}
\Line(80,50)(150,0)
\Text(152,2)[l]{$\pi^0$}
\Text(180,-2)[l]{$p_{\pi^0}$}
\end{picture}
\caption{The decay $\eta\to3\pi$  and the momenta associated with it.}
\label{figeta}
\end{minipage}
\end{figure}
This is at present a major restriction on high order ChPT predictions. It
works well for the terms not involving quark masses, OK for those with quark
masses but dominated by vectors and it is not known how well the scalar
dominated ones are predicted using this approximation.
Many calculations have been performed to order $p^6$. A recent review
is~\cite{groningen}.

A major improvement happened recently for $\pi\pi$ and $\pi K$ scattering
where the combination of $p^6$ calculations and dispersion relations
lead to a much better understanding of these systems.

\section{The decay $\eta\to3\pi$ beyond $p^4$}

One calculation which is not completed yet at $p^6$ order in ChPT is in
fact $\eta\to3\pi$. This does not mean that nothing is known
beyond order $p^4$. This section reviews that and is basically identical to
the discussion in~\cite{BG}.

The kinematics are given in terms of the via $s,t,u$ defined by
\ba
s &=&\left(p_{\pi^+}+p_{\pi^-}\right)^2=\left(p_\eta-p_{\pi^0}\right)^2
\quad\quad
t =\left(p_{\pi^-}+p_{\pi^0}\right)^2=\left(p_\eta-p_{\pi^+}\right)^2
\, \nonumber\\
u &=&\left(p_{\pi^+}+p_{\pi^0}\right)^2=\left(p_\eta-p_{\pi^-}\right)^2\,
\quad\quad
s+t+u 
= m_\eta^2+2 m_{\pi^+}^2+m_{\pi^0}^2 \equiv 3 s_0\,.
\label{defs0}
\ea
The two different amplitudes are
\ba
\langle \pi^0\pi^+\pi^-{\mbox {out}}|\eta\rangle &=& i\left(2\pi\right)^4 
\,\delta^4\left(p_\eta-p_{\pi^+}-p_{\pi^-}-p_{\pi^0}\right)
\,A(s,t,u)\,,
\nonumber\\
\langle \pi^0\pi^0\pi^0{\mbox {out}}|\eta\rangle &=& i\left(2\pi\right)^4 
\,\delta^4\left(p_\eta-p_{1}-p_{2}-p_{3}\right)
\,\overline{A}(s_1,s_2,s_3)\,.
\ea
The pions are in an $I=1$ state which means that the amplitude
is proportional to $m_u-m_d$ or $\alpha_{em}$.
The ${\cal O}(\alpha_{em})$ effect is small, but large via 
the kinematical effects of $m_{\pi^+}-m_{\pi^0}$. The photonic
decay $\eta\to\pi^+\pi^-\pi^0\gamma$ needs to be included directly.
Isospin leads to
\be
\overline{A}(s_1,s_2,s_3) = A(s_1,s_2,s_3)+A(s_2,s_3,s_1)+A(s_3,s_1,s_2)\,.
\ee

The lowest order amplitude is~\cite{Cronin67}
\be
A(s,t,u) = \frac{\dsp B_0 (m_u-m_d)}{\dsp 3 \sqrt{3} F_\pi^2}
\left\{1+\frac{3(s-s_0)}{m_\eta^2-m_\pi^2}\right\}\,,
\ee
or, with $Q^2 \equiv ({m_s^2-\hat m^2})/({m_d^2-m_u^2})$ and
$\hat m = (m_u+m_d)/2$, it becomes
\be
 A(s,t,u) =\frac{1}{Q^2} \frac{m_K^2}{m_\pi^2}(m_\pi^2-m_K^2)\,
\frac{1}{3\sqrt{3}F_\pi^2}\, M(s,t,u)\, ,
\ee
with at lowest order,
\be
\label{MLO}
M(s,t,u) = \left({\dsp 3 s-4m_\pi^2}\right)
\left({\dsp m_\eta^2-m_\pi^2}\right)\,.
\ee
That the decay rate $\Gamma\left(\eta\to3\pi\right)$ is thus
proportional to $Q^{-4}$,  allows a PRECISE measurement of $Q$.
To illustrate this we take $Q$ from the baryon mass difference,
$Q\approx 24$, and obtain at lowest order
$\Gamma(\eta\to\pi^+\pi^-\pi^0) \approx 66~\mbox{eV}\,.$
An alternative determination from $m_{K^+}^2-m_{K^0}^2\sim Q^{-2}$ gives
$Q = 20.0\pm1.5$ and leads to a lowest order prediction
$\Gamma(\eta\to\pi^+\pi^-\pi^0) \approx 140~\mbox{eV}\,.$

The $p^4$ calculation~\cite{GLeta} gives a very large enhancement
\be
\left.\left({\int dLIPS |A_2+A_4|^2}\right)\right/\left({\int dLIPS |A_2|^2}\right) = 2.4\,,
\ee
with $LIPS$ meaning Lorentz invariant phase-space.
A major source of the large effect is the large $S$-wave final state
rescattering. The $p^6$ calculation is  partially done
but has been stalled since two years.

The higher orders have been estimated via dispersion relations
that mainly include the effects of the final state rescattering.
There have been two calculations,~\cite{AL} and \cite{KWW}. They used
different methods but similar approximations. I will present a simplified
version of the analysis of~\cite{AL} here as performed in~\cite{BG}.
A more extensive description of~\cite{AL} is~\cite{Walker}.

Up to ${\cal O}(p^8)$ there are no absorptive parts from $\ell\ge2$
which allows to write{~\cite{KS,AL}
\be
M(s,t,u) =  M_0(s)+(s-u)M_1(t)+(s-t)M_1(t)
+M_2(t)+M_2(u)-\frac{2}{3}M_2(s)\,.
\ee
The $M_I$ ``roughly'' correspond to contributions with isospin 0,1,2.
The $M_I$ satisfy dispersion relation with 2 or 3 subtractions
in terms of their discontinuities
\ba
M_{0,2}(s) &=& a_{0,2}+b_{0,2} s+c_{0,2} s^2+\frac{s^3}{\pi}\,\int\,\frac{ds^\prime}{s^{\prime3}}
\,\frac{\mbox{disc} M_{0,2}(s^\prime)}{s^\prime-s-i\varepsilon}\,,
\nonumber\\
M_1(s) &=& a_1+b_1 s+\frac{s^2}{\pi}\,\int\,\frac{ds^\prime}{s^{\prime2}}
\,\frac{\mbox{disc} M_1(s^\prime)}{s^\prime-s-i\varepsilon}\,,
\label{dispMi}
\ea
The constraint on $s,t,u$ of (\ref{defs0}) implies that there are
only 4 free constants, not 8 in (\ref{dispMi}) via
\be
M(s,t,u) = a + bs + c s^2 -d (s^2 + tu)+\mbox{dispersive}\,.
\ee
The quantities
$c$ and $d$ can be determined from more convergent dispersion relations
in terms of known parameters via
\ba
\label{dispcd}
c &=& c_0+\frac{4}{3} c_2 =
\frac{1}{\pi}\,\int\,\frac{ds^\prime}{s^{\prime3}}
\,\left\{\mbox{disc} M_0(s^\prime)+\frac{4}{3}\mbox{disc} M_2(s^\prime)\right\}
\,,
\nonumber\\
d &=& -  \frac{4 L_3-1/(64\pi^2)}{F_\pi^2(m_\eta^2-m_\pi^2)}
+\frac{1}{\pi}\,\int\,\frac{ds^\prime}{s^{\prime3}}
\,\left\{s^\prime\mbox{disc} M_1(s^\prime)+\mbox{disc} M_2(s^\prime)\right\}
\ea
We now restrict the discontinuities to their respective two-body cuts and
further split them into the part coming from the forward channel,
the $M_I(s)$ part on the rhs of (\ref{dispMi2}),
and that via the $t,u$ channels,
the $\hat M_I(s)$ part on the rhs of (\ref{dispMi2}), explicitly:
\be
M_I(s) = \frac{1}{\pi}\int\,\frac{ds^\prime}{s^\prime-s-i\varepsilon}
\sin \delta_I(s) e^{-i\delta_I(s)}
\left\{M_I(s)+\hat M_I(s)\right\}\,.
\label{dispMi2}
\ee
The $\delta_I(s)$ are the $S$-wave isospin $I$ scattering phases.

The ambiguities inherent in the possible solutions
can be solved by going over to a new set
of functions that automatically include the forward cuts via
\be
\Omega_I(s) = \exp\left\{({s}/{\pi})\int {ds^\prime}/{s^\prime}\,
\,{\delta_I(s^\prime)}/({s^\prime-s-i\varepsilon})\right\}\,.
\ee
The new dispersion relations are:
\ba
\label{finaldisp}
\frac{M_0(s)}{\Omega_0(s)} &=& \alpha_0+\beta_0 s+\gamma_0 s^2
+\frac{s^2}{\pi}\int\,ds^\prime
\frac{\sin\delta_0(s^\prime)\,\hat M_0(s^\prime)}
{|\Omega_0(s^\prime)| s^{\prime2} (s^\prime-s-i\varepsilon)}\,,
\nonumber\\
\frac{M_1(s)}{\Omega_1(s)} &=& \beta_1 s
+\frac{s}{\pi}\int\,ds^\prime
\frac{\sin\delta_1(s^\prime)\,\hat M_1(s^\prime)}
{|\Omega_1(s^\prime)| s^{\prime} (s^\prime-s-i\varepsilon)}\,,
\nonumber\\
\frac{M_2(s)}{\Omega_2(s)} &=& 
\frac{s^2}{\pi}\int\,ds^\prime
\frac{\sin\delta_2(s^\prime)\,\hat M_2(s^\prime)}
{|\Omega_2(s^\prime)| s^{\prime2} (s^\prime-s-i\varepsilon)}\,.
\ea
\begin{figure}
\begin{minipage}{0.49\textwidth}
\includegraphics[width=\textwidth]{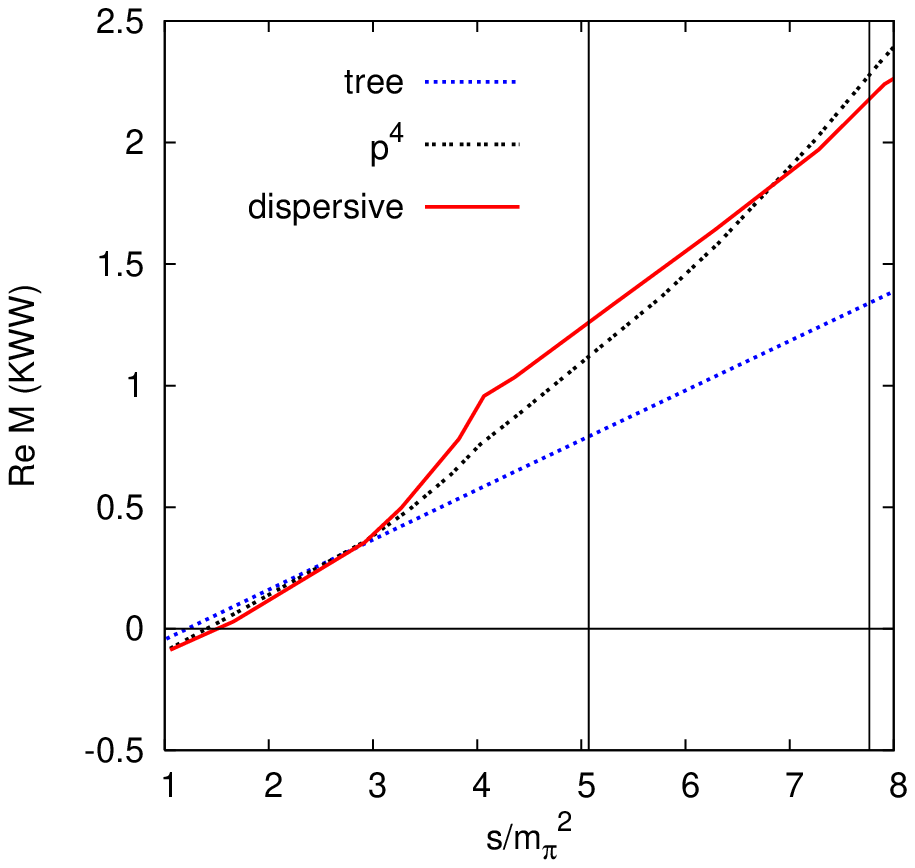}  
\end{minipage}
\begin{minipage}{0.49\textwidth}
\includegraphics[width=\textwidth]{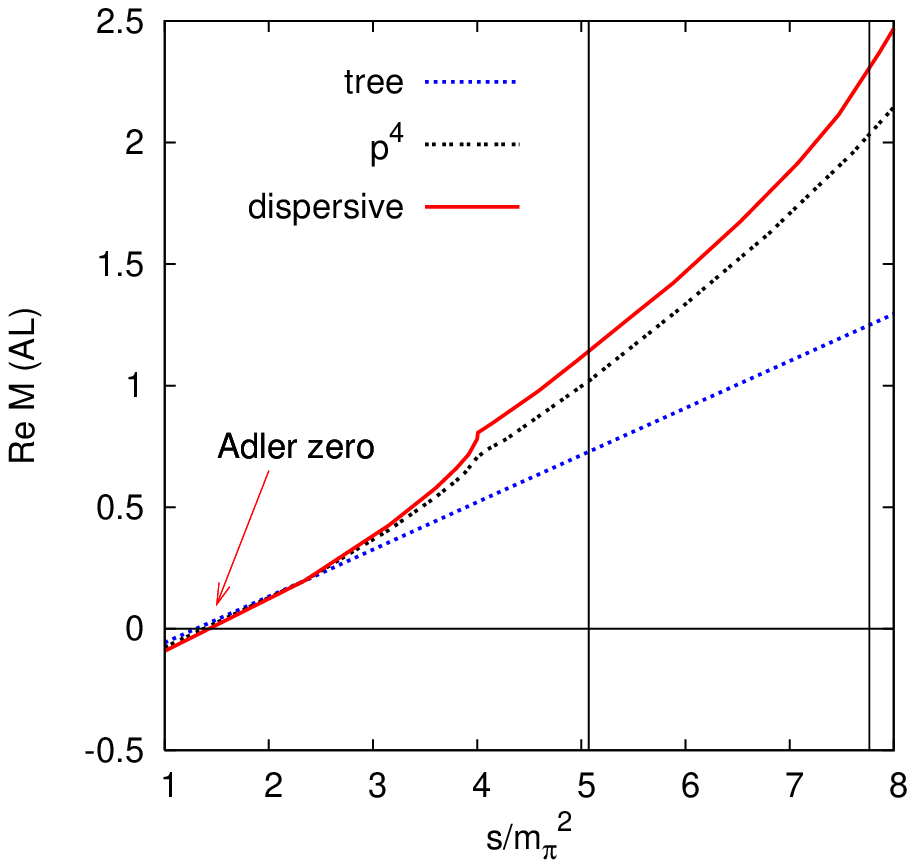}\\
\includegraphics[width=\textwidth]{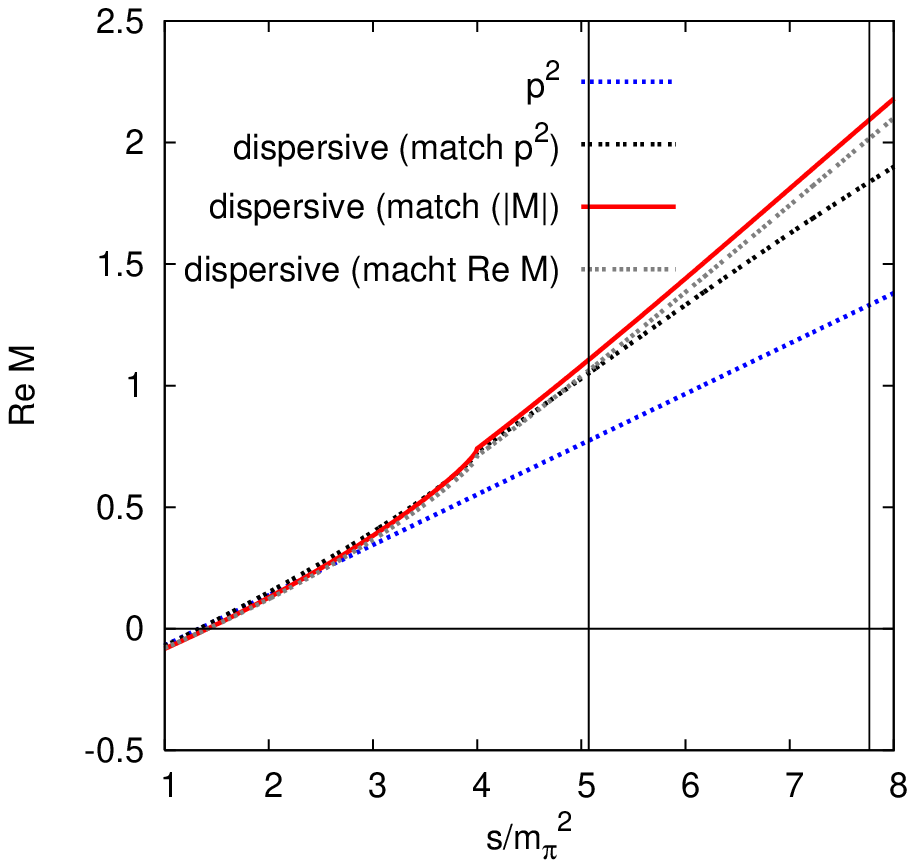}  
\end{minipage}
\caption{The real part of the $\eta\to\pi^+\pi^-\pi^0$ amplitude
plotted along the line $s=u$ with the physical
region for the decay indicated. The plots are adapted from the results
in: left~\cite{KWW}, top right~\cite{AL} and
bottom right the simplified version of~\cite{BG}. The vertical lines
note the edges of the physical region. 
}
\label{figdisp}
\end{figure}

So, we now need to find a set of $\pi\pi$ phases, $\delta_{0,1,2}(s)$,
and solve Eq.~(\ref{finaldisp}) for $M_1,M_2,M_3$.
We have to fix the four constants $\alpha_0,\beta_0,\gamma_0,\beta_1$.
Two of them,
$\gamma_0$ and $\beta_1$, can be determined from the more convergent dispersion
relations (\ref{dispcd}) with the result at $p^4$:
\be
\gamma_0\approx0\,,
\quad
\dsp\beta_1 \approx  -  \frac{4 L_3-1/(64\pi^2)}{F_\pi^2(m_\eta^2-m_\pi^2)}
\ee
The values of $\alpha_0,\beta_0$ depend on where in the 
$s,t,u$ plane matching is
done. Ref.~\cite{AL} uses that the lowest order, Eq.~(\ref{MLO}), has
an Adler zero at $s_A = 4/3 \,m_\pi^2$. This zero can move but must
remain in the neighbourhood also at higher orders. They
match the position of $s_A$ and the slope of the amplitude there
to the ${\cal O}(p^4)$ expressions.
Ref.~\cite{KWW} matches the amplitude at several places in the $s,t,u$ plane
to the ${\cal O}(p^4)$ expressions. A very simplified analysis
to understand the results
can be done by neglecting the $\hat M_I$~\cite{BG}.
The total corrections found in~\cite{AL,KWW} are very similar.
But the distributions differ significantly as can be seen from
the results along the $s=u$ line shown in Fig.~\ref{figdisp}
for the three approaches. 
The Dalitz plot distributions thus provide a check on the various input
assumptions. 
There has also been a refitting of the~\cite{KWW}
method of solving to the newer data by~\cite{Martemyanov,MS}.

The Dalitz plot distributions are parameterized by
\be
1+ ay + by^2+c x^2\quad
 \mbox {charged decay}\, ,
\quad
1+g(x^2+y^2)\quad \mbox{neutral decay}\,,
\ee
normalized at $x=y=0$. The kinematical variables are
\ba
x &=& \sqrt{3}\,\frac{T_+-T_-}{Q_\eta} = 
\frac{\sqrt{3}}{2 M_\eta Q_\eta} (u-t)\,,
\nonumber\\
y &=& \frac{3 T_0}{Q_\eta}-1\,
= \frac{3}{2 m_\eta Q_\eta}\left\{\left(m_\eta-m_{\pi^0}\right)^2-s\right\}-1
\, ,\quad
Q_\eta = m_\eta - 2 m_{\pi^+} - m_{\pi^0}
\ea
Also of interest is the ratio $
r \equiv {\Gamma(\eta\to\pi^0\pi^0\pi^0)}/{\Gamma(\eta\to\pi^+\pi^-\pi^0)}
= 1.44\pm0.04$~{\cite{PDG2004}}.
Results from the theory at $p^4$ and dispersive improvements~\cite{KWW,BG}
are given in Table~\ref{tabtheory} and the experimental results for the
neutral and charged decays in Tables~\ref{tabneutral} and \ref{tabcharged}.
You can judge the agreement yourselves.

\begin{table}
\begin{minipage}{0.65\textwidth}
\begin{tabular}{|c|cccc|}
\hline
                      & $a$     & $b$  & $c$  & $g$\\
\hline
tree                  & $-1.00$ & 0.25 & 0.00 & $0.000$\\
one-loop  & $-1.33$ & 0.42 & 0.08 & 0.03\\
dispersive (KWW) & $-1.16$ & 0.26 & 0.10 & $-0.021(7)$\\
tree dispersive       & $-1.10$ & 0.31 & 0.001& $-0.013$\\
absolute dispersive   & $-1.21$ & 0.33 & 0.04 & $-0.014$\\
\hline
\end{tabular}
\caption{Theory results for the distributions}
\label{tabtheory}
\end{minipage}
\begin{minipage}{0.34\textwidth}
\begin{tabular}{|c|c|}
\hline
                      &  $g$\\
\hline
Alde        &  $-0.044\pm0.046$\\
Crystal Barrel  & $-0.104\pm0.039$\\
Crystal Ball & $-0.062\pm0.008$\\
SND & $-0.020\pm0.023$\\
KLOE & $-0.026\pm0.014$\\
\hline
\end{tabular}
\caption{Experiment: neutral decay}
\label{tabneutral}
\end{minipage}
\end{table}

\begin{table}
\begin{center}
\begin{tabular}{|c|ccc|}
\hline
                      & $a$     & $b$  & $c$ \\
\hline
Layter   & $-1.08\pm0.14$ & $0.034\pm0.027$&$0.046\pm0.031$  \\
Gormley & $-1.17\pm0.02$ & $0.21\pm0.03$ & $0.06\pm0.04$ \\
Crystal Barrel 
                      & $-0.94\pm0.15$ & $0.11\pm0.27$ &  \\
Crystal Barrel        
             & $-1.22\pm0.07$ & $0.22\pm0.11$ & 0.06 fixed \\
KLOE & $-1.072\pm0.009$ & $0.117\pm0.008$ & $0.047\pm0.008$  \\
\hline
\end{tabular}
\end{center}
\caption{Experimental results for the charged decay}
\label{tabcharged}
\end{table}

\section{Chiral Lagrangians and $\eta'$}

To treat the $\eta'$ we have to go back and look at the mechanism that
gave the $\eta'$ its mass. This is the $U(1)_A$ anomaly
as discussed earlier. There exists a limit in QCD where the anomaly is not
there, namely when the number of colours, $N_c$, is not kept at
three but sent to infinity keeping $N_c\alpha_S$ constant~\cite{largeN}.
In this limit many aspects of QCD simplify, see e.g.~\cite{Manohar} for
an overview and introduction. The $\eta'$ becomes
a Goldstone Boson and it can be introduced in the Lagrangian.
This was done by Veneziano, DiVecchia, Witten, Schechter
and others~\cite{etap1,etap2,etap3,etap4,etap5}.
One treats the parameter $\theta$ as an external field and adds the
singlet degree of freedom $\phi_0$ to the Goldstone boson matrix $U$
as follows:
\be
\tilde\phi=\theta+\frac{\sqrt{2}\phi_0}{F},\quad
U = e^{\dsp i\sqrt{2}\phi_0/F_0}e^{\dsp i\sqrt{2}M/F}.
\ee
Under a general $U(3)_L\times U(3)_R$ transformation $\tilde\phi$
is invariant and $U\to g_R U g_L^\dagger$. The Lagrangian must then be
constructed being invariant under the full $U(3)_L\times U(3)_R$.
The transformation of $\theta$ as an external field assures that the
effect of the anomaly is correctly accounted for. This approach has two
problems. It is not clear whether it is a convergent procedure with a
derivative expansion for the $\eta'$ and the number of free parameters
is very large. In fact, since $\tilde\phi$ is invariant one can add free
functions $F_i(\tilde\phi)$ instead of all the usual parameters of ChPT
as well as a series of extra terms. To lowest order there are 5 such
functions~\cite{GL2} but at next order there are 57~\cite{herrera}.

Luckily, if the large $N_c$ counting itself is included into the
power counting things become simpler. By looking at
\be
\partial_\mu A^{0\mu}[{\cal O}(N_c)] &=& m_q P [{\cal O}(N_c)]
~+~ \omega[{\cal O}(1)]\,,
\ee
we see that we can take $\omega$ as a perturbation and treat $\tilde\phi$
as a quantity of order $1/\sqrt{N_c}$. The leading part is then the
usual chiral Lagrangian but with the nonet $U$ and a mass term for the
singlet added as discussed in~\cite{etap1,etap2,etap3,etap4}.
Treating $\omega$ as a perturbation is the basis of {\em all} the large $N_c$
chiral Lagrangian predictions for the $\eta'$.

\section{The decays $\eta'\to\eta\pi\pi$ and $\eta'\to3\pi$}

We can now use the methods discussed in the previous section but some problems
remain. There are large $\pi\pi$ rescatterings
possible in the $S$-wave channel, which are $1/N_c$ suppressed but sizable.
In other words, how do we deal with the ``$\sigma$''?
The other problem is that $\rho$ and $\omega$ are present in the
final states. One thus obviously needs to go beyond ChPT.
This also means that experiment in these decays will provide us with needed
clues on how we can go beyond ChPT. Several attempts at resummation
exist. One example is the work by Borasoy and
collaborators~\cite{Borasoy,Borasoy2} and the unitary extension of
chiral perturbation theory. The latter can be traced back from~\cite{Oset}.

The lowest order Lagrangian has three terms
\be
{\cal L}&=&
\frac{F^2}{4}\langle D_\mu U D^\mu U^\dagger\rangle ~[(a)]~
+\frac{F^2}{4}\langle \chi U^\dagger + U \chi^\dagger\rangle ~[(b)]~
-\frac{1}{2} m_0^2 \phi_0^2\,.
\label{etapLO}
\ee
The two decays are very different in their origin.
The amplitude for
$\eta'\to\eta\pi\pi$ comes from the term (a) in Eq.~(\ref{etapLO})
with
\be
 A(\eta'\to\eta\pi\pi) = 
\frac{m_\pi^2}{6 F}\left(s\sqrt{2}\cos(2\theta)-\sin(2\theta)\right)
\ee
while
$\eta'\to\pi\pi\pi$ is produced from term (b) in Eq.~(\ref{etapLO})
with an amplitude
\be
A(\eta'\to\pi\pi\pi)\sim \frac{m_u-m_d}{F_\pi}\,.
\ee
The predictions from these are shown in Table~\ref{tabetap}
and compared with the experimental values.
\begin{table}
\begin{center}
\begin{tabular}{|c|cc|cc|}
\hline
 & $\eta'\to\eta\pi^0\pi^0$ & 
$\eta'\to\eta\pi^+\pi^-$ & 
$\eta'\to\pi^0\pi^0\pi^0$ & 
$\eta'\to\pi^0\pi^+\pi^-$ \\
\hline
 & 1.0 keV &  1.9 keV &  455 eV &  405 eV\\
Exp & $42\pm6$ keV & $89\pm10$ keV & $311\pm77$ eV & $\le 1005$ eV\\
\hline
\end{tabular}
\end{center}
\caption{The prediction from the lowest order chiral Lagrangian for
$\eta'$ decays and their experimental values.}
\label{tabetap}
\end{table}
The two types of decays are quite different. The decays $\eta'\to\eta\pi\pi$
are very much off the predictions while the $\eta'\to3\pi$ agree reasonably
well.  In the former decay, no isospin breaking is needed. The factor of
$m_\pi^2$ which suppresses the decay in fact disappears at higher orders.
The next order in $1/N_c$ allows for terms like
$\eta'\eta\partial_\mu\pi\partial^\mu\pi$ which remove this suppression.
That alone allows one to fit the rate but not the slope. We therefore need to
study the distributions in phase space to uncover the type of effects
that are important. The precision of the branching ratios also needs
improvement to check the factor of 2 prediction from isospin.
The $3\pi$ decay is isospin violating and always needs an overall factor
of $m_u-m_d$. We thus do not expect as dramatic effects compared to lowest
order but to check good quantitative agreement needs better experimental
precision.

\section{Anomalies}

A major topic in $\eta$ and $\eta'$ decays is the study of the nonabelian
anomaly~\cite{Bardeen} and its effective Lagrangian, the Witten-Wess-Zumino
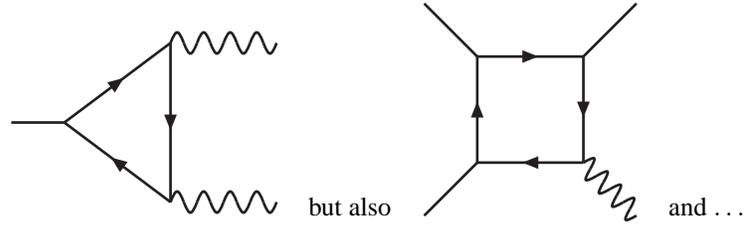
\begin{figure}
\begin {center}
\unitlength=1pt
\begin{picture}(120,90)(0,15)
\SetScale{1.0}
\SetWidth{1.}
\Line(10,50)(30,50)
\ArrowLine(30,50)(70,80)
\ArrowLine(70,80)(70,20)
\ArrowLine(70,20)(30,50)
\Photon(70,80)(110,80){4}{4}
\Photon(70,20)(110,20){4}{4}
\end{picture}
but also
\unitlength=1pt
\begin{picture}(100,100)(-10,0)
\SetScale{1.0}
\SetWidth{1.}
\Line(0,0)(20,20)
\Line(0,80)(20,60)
\Line(60,60)(80,80)
\ArrowLine(20,20)(20,60)
\ArrowLine(20,60)(60,60)
\ArrowLine(60,60)(60,20)
\ArrowLine(60,20)(20,20)
\Photon(60,20)(80,0){4}{4}
\end{picture}
and $\ldots$
\end{center}
\caption{Some of the graphs for the anomaly. Now with external photons
or weak bosons rather than gluons.}
\label{anomalygraphs}
\end{figure}
term~\cite{Wess,Witten6}. An obviously incomplete list of processes and
experiments is given in Table~\ref{tabanomaly}.

The ones labeled $g-2$ are interesting also for the muon anomalous magnetic
moment. They show up as subparts of the light-by-light hadronic contribution
there, as shown in Fig.~\ref{figgm2}.
The order $p^6$ Lagrangian is known~\cite{anop6}.
An earlier review for anomalous $\eta$
process is~\cite{anoreview}.

\section{Conclusions}

There is a rich field of physics to be explored in $\eta$ and $\eta'$ decays.
Some of these have been discussed in this talk. We look forward to
find out more from WASA, KLOE and the other experiments presented
at this meeting. Let me conclude by giving simply a list of topics/questions.
\begin{itemize}
\itemsep0cm
\item Precision physics: { $Q$ from $\eta\to3\pi$}.
\item Understanding physics: {} $\pi^0$ versus $\eta$ versus
$\eta'$.
\item All channels: do the flavour singlet degrees of freedom 
differ significantly from the nonsinglet?
\item Glue is important for $\eta'$ in its mass. Can we detect it
 also in other places or is the rest \emph{merely}
a problem of final state interactions?
\item requires getting at the mechanisms behind $\eta,\eta'$ decays.
\item High quality distributions are a must.
\end{itemize}

\begin{table}
\begin{center}
\begin{tabular}{|cc|ccc|}
\hline
$\pi^0\to\gamma\gamma$ & Primex, $e^+e^-$ &
$\eta'\to\pi^+\pi^-\gamma^{(*)}$ & WASA &\\
$\eta\to\gamma\gamma$ & Primex, $e^+e^-$ &
$\eta\to\gamma^{(*)}\gamma^{(*)}$ & WASA,KLOE, CLEOc & g-2\\
$\eta'\to\gamma\gamma$ & $e^+e^-$ &
$\eta'\to\gamma^{(*)}\gamma^{(*)}$ & WASA,KLOE, CLEOc & g-2\\
$\gamma\pi^0\pi^+\pi^-$ & Primakoff &
$\eta'\to\rho^0\gamma$ & WASA & \\
$\eta\to\pi^+\pi^-\gamma^{(*)}$ & WASA,KLOE & & & \\
\hline
\end{tabular}
\end{center}
\caption{Some anomalous processes and the experiments where we expect
improvements. The symbol
$\gamma^{(*)}$ stands for (if allowed) $\gamma$, $e^+e^-$, $\mu^+\mu^-$
or an off-shell photon in tagged $\gamma\gamma$ collisions.}
\label{tabanomaly}
\end{table}
\begin{figure}
\hskip1cm\includegraphics[width=5cm]{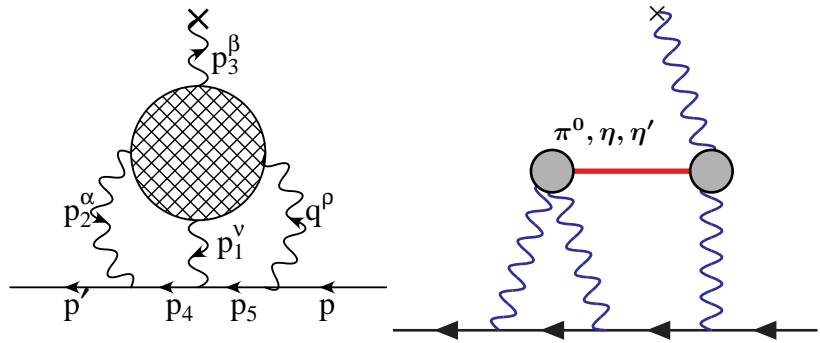}
\unitlength=2pt
\begin{picture}(80,70)
\SetScale{2.0}
\SetWidth{0.5}
\ArrowLine(20,0)(0,0)
\ArrowLine(40,0)(20,0)
\ArrowLine(60,0)(40,0)
\ArrowLine(80,0)(60,0)
\Text(50,60)[c]{\boldmath$\times$}
\SetColor{Blue}
\Photon(20,0)(30,30){2}{6}
\Photon(40,0)(30,30){2}{6}
\Photon(60,0)(60,30){2}{6}
\Photon(60,30)(50,60){2}{6}
\SetColor{Red}
\SetWidth{1.}
\Line(30,30)(60,30)
\SetWidth{0.5}
\Text(40,35)[b]{\boldmath$\pi^0,\eta,\eta'$}
\SetColor{Black}
\GCirc(30,30){4}{0.7}
\GCirc(60,30){4}{0.7}
\end{picture}
\caption{The light-by-light hadronic contribution to the muon $g-2$ and
a subprocess with anomalous vertices.}
\label{figgm2}
\end{figure}

\begin{ack}
I thank the organizers for a very cordial and exciting meeting.
Supported by the Swedish Research Council,
the EU TMR 
HPRN-CT-2002-00311  (EURIDICE) and
the EU-Research Infrastructure
Activity RII3-CT-2004-506078 (HadronPhysics).
\end{ack}

\end{document}